\documentclass[usenatbib,useAMS,twocolumn]{mn2e}

\usepackage{subfigure}
\usepackage{graphicx}
\usepackage[latin1]{inputenc}
\usepackage{psfig,epsfig}
\usepackage{amssymb}
\usepackage{amsmath}
\bibpunct{(}{)}{;}{a}{}{,}
\newcommand\gb[1] {   \mbox{\boldmath{$#1$}}  }

\newcommand\del{\gb{\nabla}}
\newcommand\bdot{\gb{\cdot}}
\newcommand\btimes{\gb{\times}}
\newcommand\vv{\gb{v}}
\newcommand\uu{\gb{u}}
\newcommand\B{\gb{B}}
\newcommand\Bb{\gb{b}}
\newcommand\by{\bar{y}}
\newcommand\byc{\bar{y}_{\rm c}}
\newcommand\byp{\bar{y}_{\rm p}}
\newcommand\yc{y_{\rm c}}
\newcommand\yp{y_{\rm p}}
\newcommand\bT{\bar{T}}

\newcommand\bQ{\gb{\sigma}}

\newcommand\bdiv[1]{\del \bdot #1}

\newcommand\hy{\gb{\hat{y}}}
\newcommand\hz{\gb{\hat{z}}}

\newcommand{\dd}[2]{\frac{{\rm d} #1}{{\rm d} #2}}
\newcommand{\Dd}[2]{\frac{{\rm D} #1}{{\rm D} #2}}
\newcommand{\dpart}[2]{\frac{\partial #1}{\partial #2}}

\newcommand{\bma}[1]{ \left[ \begin{array}{#1}}
\newcommand{\ema}{\end{array} \right] }

\def\beq{ \begin{equation} }
\def\eeq{ \end{equation} }
\def\ltsim{\mathrel{\spose{\lower.5ex\hbox{$\mathchar"218$}}
\raise.4ex\hbox{$\mathchar"13C$}}}
\def\gtsim{\mathrel{\spose{\lower.5ex\hbox{$\mathchar"218$}}
\raise.4ex\hbox{$\,\rangle $}}}
\title[Local simulations vs reduced models] { A comparison of local
  simulations and reduced models of MRI-induced turbulence}
\author[Pierre Lesaffre, Steven A. Balbus  \& Henrik Latter]
{Pierre Lesaffre$^{1}$\thanks{Email:
pierre.lesaffre@lra.ens.fr},
Steven A. Balbus$^{1,2}$ and
Henrik Latter$^{1}$  \\
$^{1}$ Laboratoire de
Radioastronomie, 24 rue Lhomond, 75231 PARIS Cedex 05, France\\
$^{2}$ Adjunct Professor, Dept. of Astronomy, University of Virginia, Charlottesville V1 22903}
\begin{document}
\date{Received}
\maketitle
\begin{abstract}

 We run mean-field  shearing-box numerical  simulations with a
 temperature-dependent resistivity and compare them to a reduced
 dynamical model.  Our simulations reveal the co-existence of two
 quasi-steady states, a `quiet' state and an `active'
 turbulent state, confirming the predictions of the reduced model.
 The initial conditions determine on which state the simulation
 ultimately settles.  The active state is strongly influenced by the
 geometry of the computational box and the thermal properties of the
 gas. Cubic domains support permanent channel flows, bar-shaped
 domains exhibit eruptive behaviour, and horizontal slabs give rise to
 infrequent channels.  Meanwhile, longer cooling time scales lead to
 higher saturation amplitudes. 
\end{abstract}
\begin{keywords}
MHD --
turbulence --
methods: numerical --
accretion, accretion discs --
planetary systems: protoplanetary discs
\end{keywords}
\section{Introduction}
The observed accretion luminosity of astrophysical disks demands that
  angular momentum be transported at a rate far in excess of that
  possible by molecular viscosity in a laminar flow \citep{SS73}. The
  magneto-rotational instability (MRI) \citep{BH91} provides a more
  efficient mechanism: the MRI generates turbulent motions that are
  strongly correlated and which significantly increase the effective
  flux of angular momentum.  It is hence regarded  the most
  promising candidate responsible for 
the observed anomalous transport in
  magnetised disks. Our main observational probes of disks exploit
  their radiation properties, which in turn strongly depend on their
  thermal and chemical properties. So in order to connect observations
  with the dynamical state of accretion disks we must better
  understand the relationship between MRI turbulence and the chemistry
  and thermodynamics of accretion disks.  The goal of this paper is to
  make progress to this end.

 We shall focus principally on the role of the thermodynamics and in
 particular, resistive dissipation.  From the outset we assume a
 generic model of radiative loss and fast chemistry, so that the
 resistivity $\eta$ depends only on temperature. The latter
 temperature dependence, however, gives rise to an interesting
 feedback on the turbulent dynamics. Fluctuations in temperature will
 not only alter the resistive scale, and hence the turbulent cascade
 of energy \citep[see][]{F07II,LL07}, it will also influence the linear MRI
 modes which drive the turbulence itself in regions where the
 resistivity is sufficiently high. 

In order to clarify the intriguing relationship between the disk
thermodynamics and the MRI-induced turbulent dynamics, \cite{BL08}
devised a simple 2 variable  dynamical system, which describes the
evolution of the temperature, on the one hand, and the magnitude of the
turbulence, on the other. The model incorporates both the dynamics of
the MRI,  heating (coupling the turbulent fluctuations to
the temperature), radiative cooling  and a temperature
  dependent resistivity $\eta$ (coupling the temperature to the
  turbulent fluctuations).  Under very general circumstances the simple
system possesses two stable equilibria (or fixed points).  One fixed
point corresponds to a cold flow --- the `quiet' state, in
which the resistivity is high, the MRI shuts off, and the temperature
is set by radiative balance. The other fixed point corresponds to a
hot turbulent flow --- the `active' state, in which dissipative
heating balances the radiative gains and losses.  Which state the
system selects ultimately depends on the initial conditions.  
Accretion disks are potentially susceptible to this kind of phase
separation. They need only experience a broad enough range of
temperatures so that both active and quiet states coexist.
For example, protoplanetary disks \citep{Fr02} and
dwarf-novae disks \citep{GM98} may possibly experience large enough
resistivities to have a direct influence on outer-scale dynamics.
 
In this paper we show that this framework
can be used to interpret more sophisticated models of disks, namely
numerical simulations of shearing boxes. 
 We restrict ourselves here to
unstratified shearing boxes with a net mean vertical magnetic flux. We
also investigate the effect of changing the aspect ratio of the
computational box, as this seems to strongly influence the qualitative
results.  We first describe the reduced dynamical model (Section
2). The details of the numerical setup are presented in the Appendix.

We then present our results, first comparing the simulations in cubic
domains with the reduced model (Section 3), then discussing the aspect
ratio dependence of the simulations' results (Section 4).  In Section 5 we draw our conclusions.

\section{Shearing box reduced model}

\subsection{Governing equations for the 3D simulations}
\label{govequ}
Our simulations use a standard shearing-box setup. The frame of reference
rotates at circular angular velocity $\Omega$.  Radial, azimuthal and
vertical directions are labelled by local Cartesian coordinates $x$,
$y$ and $z$.  The origin of the frame follows an unperturbed fluid
element moving in a circular orbit.  The radial logarithmic derivative
of $\Omega$ is $$q=\frac12\dd{\Omega^2 }{\ln{R}}|_{x=0}$$ and
characterises the local shear. It takes the value $q=3/2$ for a Keplerian shear and we adopt this value for the remainder of the paper. This shearing sheet system is then
supplemented with shearing box boundary conditions at the edges of the
computational domain \citep{HGB95}.

The fundamental dynamical equations in this rotating frame are
the mass continuity equation, 
\beq
\label{continuity}
\dpart{\rho}{t}+\bdiv{(\rho\vv)}=0 ,
\eeq
where $\rho$ is the mass density of the gas and $\vv$ is its velocity,
and the Navier-Stokes equation with a kinematic viscosity $\nu$,
$$
\dpart{\vv}{t}+(\vv\bdot\del)\vv+2\Omega\gb{\hz \times v}+\del \Phi
+\frac1{\rho}\del(p+\frac{B^2}{8\pi})-\frac1{4\pi\rho}(\B\bdot\del)\B
$$
\beq
\label{Euler}
=\frac1{\rho} \bdiv(\rho \nu \bQ) ,
\eeq
 where $p$ is the thermal pressure, $\B$ is the magnetic field,
 $\Phi=q\Omega x^2$ is the tidal potential and
 $\sigma_{ij}=\frac12(\partial_i v_j+\partial_j v_i)-\frac13\partial_k v_k\delta_{ij}$
is the stress tensor. Vertical gravity is neglected.
The induction equation is
\beq
\label{induction}
\dpart{\B}{t}=\del\btimes(\vv\btimes\B-\eta(T)\gb{J})
\eeq
where $\gb{J} \equiv \del\btimes\B$ and $\eta(T)$ is the resistivity (a decreasing function of the temperature, see equation  (\ref{etat})).

We run both isothermal and non-isothermal simulations.
In the non-isothermal case, we adopt an ideal gas equation of state with adiabatic index
$\gamma$ so that the internal energy is
\beq
e=\frac1{\gamma-1}p
\eeq
and the equation for the Lagrangian derivative of the entropy reads
\beq
\label{energy}
e\Dd{\ln (p \rho^{-\gamma})}{t} = \eta J^2+\rho\nu \bQ\gb{:\nabla v}-
\rho\Lambda(T)- \del\bdot(\chi \rho \del T) 
\eeq 
 where $\Lambda(T)=aT+b$ is a linear net cooling function (including radiative
gains and losses), $\chi$ is the thermal diffusion coefficient and $T$
is the specific energy defined as $e/\rho$ (it is hence proportional to the temperature).  
The adiabatic index $\gamma$
has been fixed to the standard value of $5/3$ for the present study.
More details on the thermal properties of our simulations are found in
 Appendix
\ref{thermal}.

   Equations (\ref{continuity}) to
(\ref{energy}) form the governing system of which we seek approximate
numerical solutions.
The steady state solution consists of the state of homogeneous density
and linear shear: $\rho=\rho_0$ and $\vv= A\, x\,\hy \equiv \vv_0 $
and we take a background magnetic field constant and vertical, $\B=
B_0\,\hz$. We denote
the perturbation velocity and magnetic field as $\uu=\vv-\vv_0$ and
$\Bb=\B-B_0\,\hz$. We use a version of the \small{ZEUS3D} code \citep{ZI,ZII} suited
to our needs (Appendix \ref{setup} contains the full details of our numerical method).

\subsection{Simple dynamical model}
  The reduced model of \cite{BL08} may be represented as:
  \beq \label{e1} \dd{y}{t}=\sigma(T) y-Ay^n \eeq for the evolution of
  a generic turbulent amplitude $y$ in the gas and \beq \label{e2}
  \dd{T}{t}=Wy^2-\Lambda(T) \eeq for its temperature $T$.  The time $t$ is
  normalised to the orbital timescale $1/\Omega$.  The linear
  growth/damping rate of $y$ is denoted by $\sigma$. It depends on the
  temperature $T$ via the resistivity $\eta(T)$. We follow \cite{BL08}
  and use \beq \sigma(T)=\sigma_*[1-\eta(T)/\eta_*] \eeq where
  $\sigma_*$ is the ideal MRI rate of growth at the computational box
  wavelength (for $\eta=0$) and $\eta_*$ is the critical resistivity
  for MRI at that wavelength (i.e. when $\eta=\eta_*$, the fastest
  growing mode in the computational box is marginally stable).  The
  parameters $A$ and $n$ account for the non-linear saturation
  dynamics of the gas. The  function $\Lambda(T)$ denotes the
  radiative losses.  And the constant $W$ accounts for the dissipation of the turbulent kinetic energy at small scales. For stationary turbulence, 
  this is equivalent to the energy input at large scales.

But   the comparison is quantitatively better if we improve equation
  \eqref{e2} in our dynamical model to account for the finite time
  over which energy is transferred from large to small scales. We hence
  adopt the new system of reduced equations 
\beq
\label{red2a}
\dd{y}{t}=\sigma(T) y-Ay^n \eeq
and
\beq
\label{red2b}
  \dd{T}{t}=Wy^2-\dd{y^2}{t}-\Lambda(T) \mbox{.}
\eeq
The additional term  d$y^2/$d$t$ with a time-derivative does not
change the fixed points of the dynamical system and only slightly modifies their
stability properties.

\subsection{Phase diagram of the reduced model}

\cite{BL08} have shown that, under very general circumstances (namely:
$\eta(T)$ decreasing and $\Lambda(T)$ increasing to infinity), the above
dynamical system possesses the same phase diagram (illustrated on Fig. 1). It exhibits
three fixed points. We have already discussed the two stable fixed points `quiet' and `active' states in the introduction. We label them respectively `Q' and `A'.  

The third fixed point is
always unstable and occurs for intermediate temperatures: following
\cite{BL08} we label it `I'. An initial state $(y,T)$ which finds itself
near the coordinates of the point `I' ends up at `A' or `Q'
depending on its position relative to the separatrices of the point `I'. 
These two separatrices define four quadrants in the $(T,y)$ phase space.
 A cloud of various
points evolving around point `I'  splits into
two cloudlets which end up one on the quiet region and one on the
active region.
In effect, this system describes the phase separation between active
and quiet regions.

\subsection{Reduced variables}
 In order to compare the numerical solution with the reduced model, 
we must extract from the simulations two quantities analogous to $T$ and $y$
 at
each time step. We label these numerical variables $\bT$ and $\by$.

  We start with the total energy conservation \citep[see equation (8)
    of][for example]{HGB95} where we neglect the viscous stresses: 
\beq \dd{}{t}\langle\, e+m\,\rangle =\frac32\langle\, W_{xy}\,\rangle -\langle\, \rho
  \Lambda(T)\,\rangle  \mbox{.}\eeq 
Here the angles denote volumic averages over the computational domain, $e$ is the internal energy, $m=\frac12\rho
v^2+\frac1{8\pi}b^2+\rho\Phi$ is the total mechanical energy
(including the tidal potential $\Phi$) and $W_{xy}=\rho u_x u_y -
\frac1{4\pi}b_x b_y$ is the total radial-azimuthal stress tensor of
the perturbation. The variables $\rho$, $\vv$, $\uu$ and $\Bb$ are
respectively the mass density, the total velocity, the perturbed
velocity and magnetic field (see Section \ref{govequ}).
We normalise the time with the orbital timescale  so
that $\Omega=1$ in these time units.  In addition, we use a linear
cooling function $\Lambda(T)=aT+b$ with $a$ and $b$ constant
parameters and we write:
  \beq
\label{red2}
\dd{\langle\, e\,\rangle }{t}=\frac32\langle\, W_{xy}\,\rangle -\dd{\langle\, m-m_0\,\rangle }{t}-\Lambda(\langle\, e\,\rangle ) \mbox{.}
\eeq
 We offset the mechanical energy $m$ by its value $m_0$ (constant in time) on the equilibrium velocity profile.
 If we further neglect the potential energy in the mechanical energy, then 
\beq \langle\, m-m_0\,\rangle \simeq \langle\, \frac12\rho u^2+\frac1{8\pi}b^2\,\rangle \mbox{.}\eeq
  Hence both $\langle\, W_{xy}\,\rangle$ and $\langle\, m-m_0\,\rangle$ are quadratic in the perturbation amplitude.
 The comparison of equation \eqref{red2} with equation \eqref{e2}
leads us to use \beq \bT=\langle\, e\,\rangle \eeq as the reduced variable for
temperature and \beq \by^2=\langle\, \frac12\rho u^2+\frac1{8\pi}b^2\,\rangle \eeq as the reduced variable
for the turbulent amplitude.

\section{Results for cubic domains}

We select four initial conditions in each of the four quadrants that
produce $(\bT,\by)$ simulation trajectories qualitatively similar to
the $(T,y)$ trajectories of the reduced model (thick coloured lines
 in Fig. \ref{phase}).

\subsection{Parameters}
In
order to make the comparison more quantitative, we need to specify the
parameters $A$, $n$ and $W$. We use the estimated positions of the points `I' 
and `A' in the phase space $(\bT,\by)$ in order to constrain the set of
 parameters that yields the best agreement.
 Indeed, the 
coordinates $(T_{\rm I},y_{\rm I})$ for the critical point `I' completely specify
the parameters $A$ and $W$ for a given value of $n$:
\beq A=\sigma(T_I) y_I^{1-n} \eeq 
and \beq
W=\frac{\Lambda(T_{\rm I})}{y_{\rm I}^2} \mbox{.}  \eeq
  
The point `I' is located at the crossing point between the
separatrices, which are the asymptotes of the trajectories at early
and late times. We estimate the position of the point `I' for the simulations  from the point at
which the red and green trajectories diverge (see Fig. \ref{smallphase}).

In fact, the equations above also hold with $A$ indices : the active
saturation point can also be used to fit the parameters.  We find that $A=0.1$, $n=2$
and $W=1.2$ are values which satisfy both the conditions for points `I'
and `A'. 

\subsection{Phase diagram comparison}
We plot the simulation trajectories
in Fig.s \ref{smallphase} and \ref{bigphase}.  These should be compared with
the trajectories in the first Fig. 1. As is plain, the qualitative agreement between
the two methods is excellent. Particularly remarkable is the quantitative
agreement near the critical point.
 This demonstrates that the relative values of thermal and turbulent
time scales are accurately reproduced by the reduced model (the phase
space trajectories are governed by the ratio between d$y/$d$t$ and d$T/$d$t$)).  Fig.
\ref{bigphase} also shows that the reduced model can describe the
behaviour of the 3D system in a consistent way over a great range of
magnitudes for $y$.  The small value for the parameter $A$ actually
helps the linear approximation to hold until larger amplitudes for
$y$. The fastest linear modes are then likely to grow to large
amplitudes relative to the other modes, which makes this system
suitable to parasitic analyses such as in \cite{GX94} or \cite{LLB09}.

Our reduced model is in effect a 2 variable projection of a system
with many more degrees of freedom. But surprisingly few details show
the limits of the reduced model: only some trajectories exhibit
self-crossing behaviour, for example when the sytem wanders near
the saturation point `A'.

\begin{figure}
\centering
\psfig{file=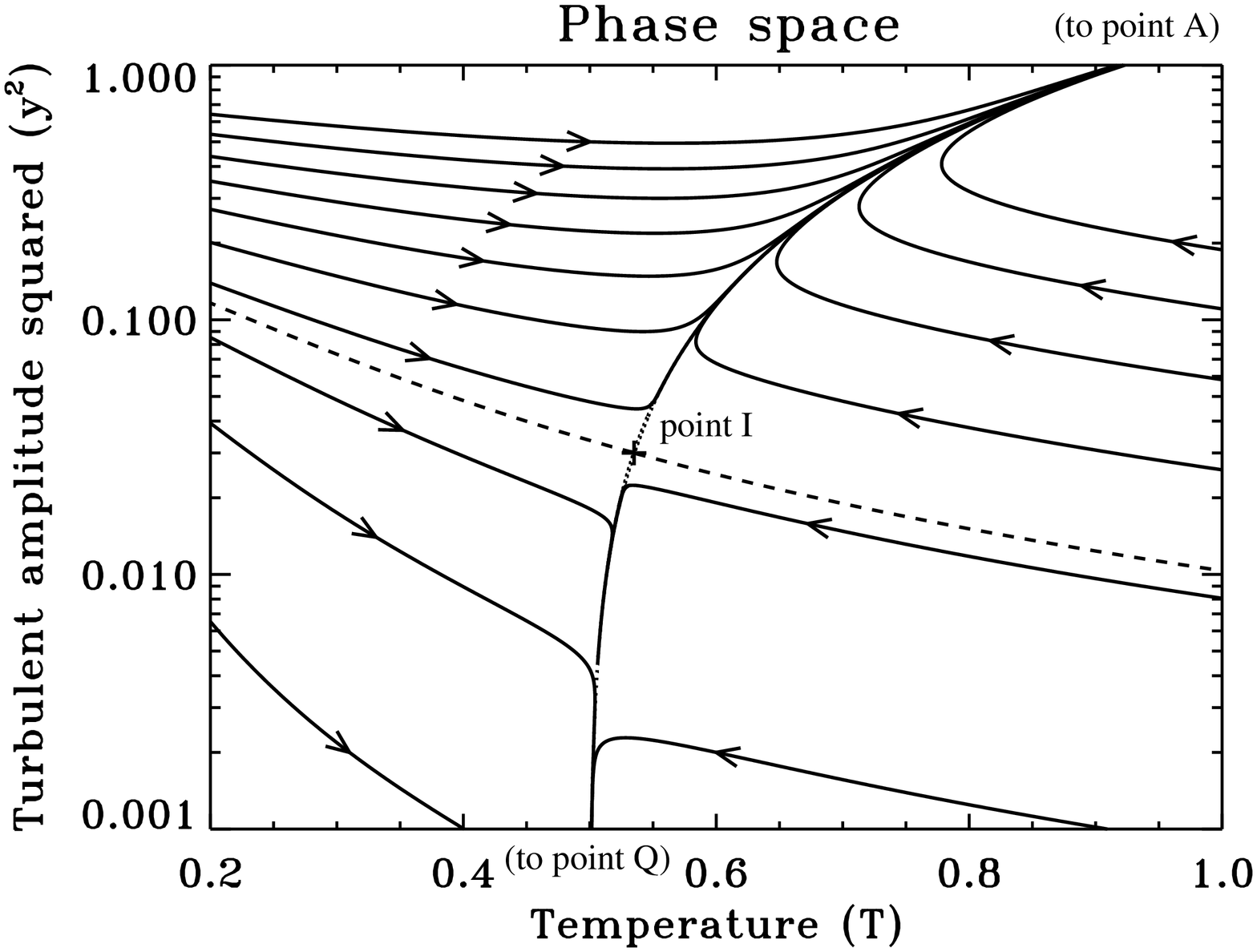,width=6cm,angle=-90}
\label{redphase}
\caption{Phase diagram of the reduced model equations 
(\ref{red2a}) and (\ref{red2b}) with parameters $A=0.1$, $n=2$ and
  $W=1.2$. The separatrices of point `I' are
shown as dotted and dashed lines.
} 
\end{figure}

\begin{figure}
\centering
\subfigure[Close up on the point `I'.]{
\psfig{file=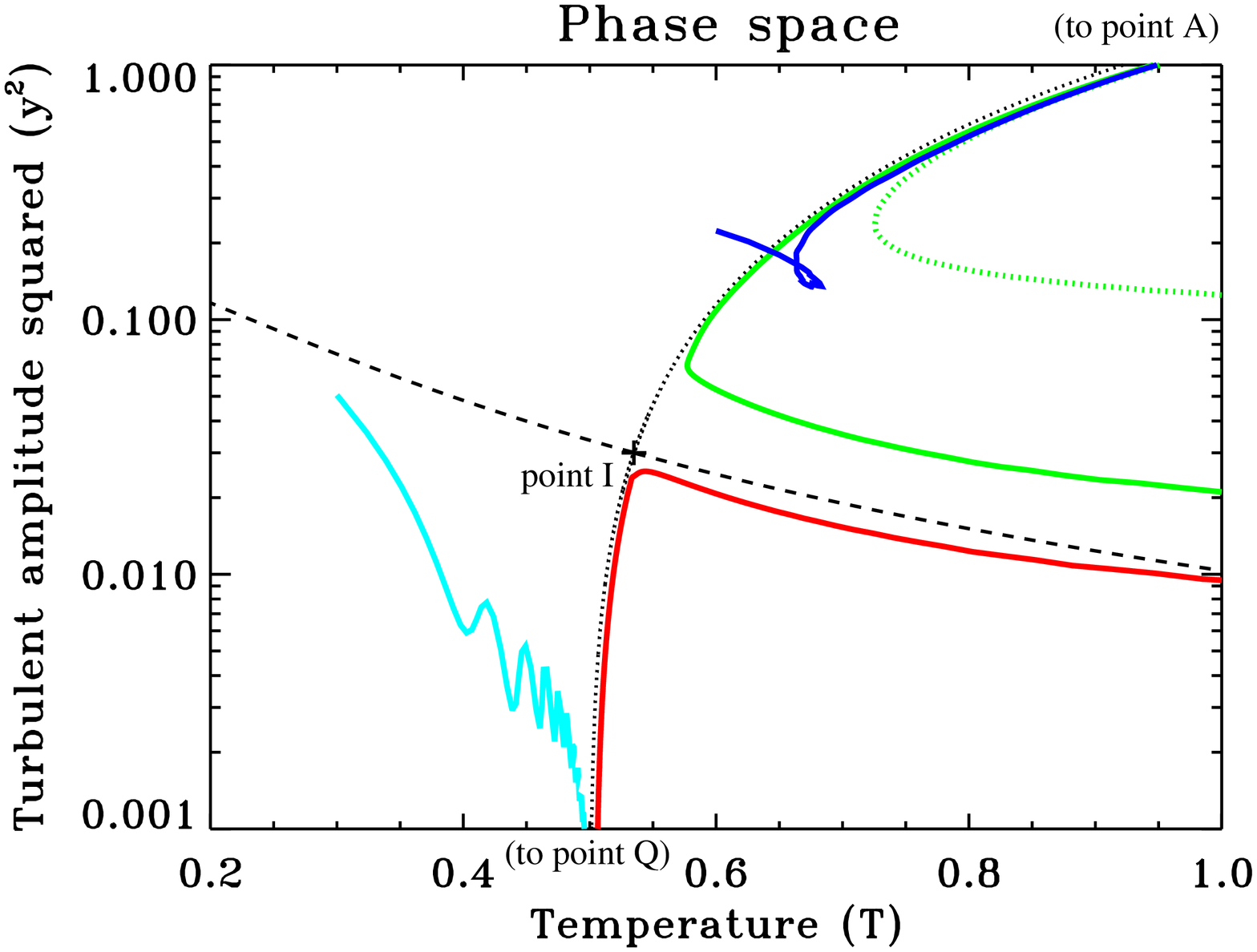,width=6cm,angle=-90}
\label{smallphase}
} 
\subfigure[Global view of the phase diagram.]{
\psfig{file=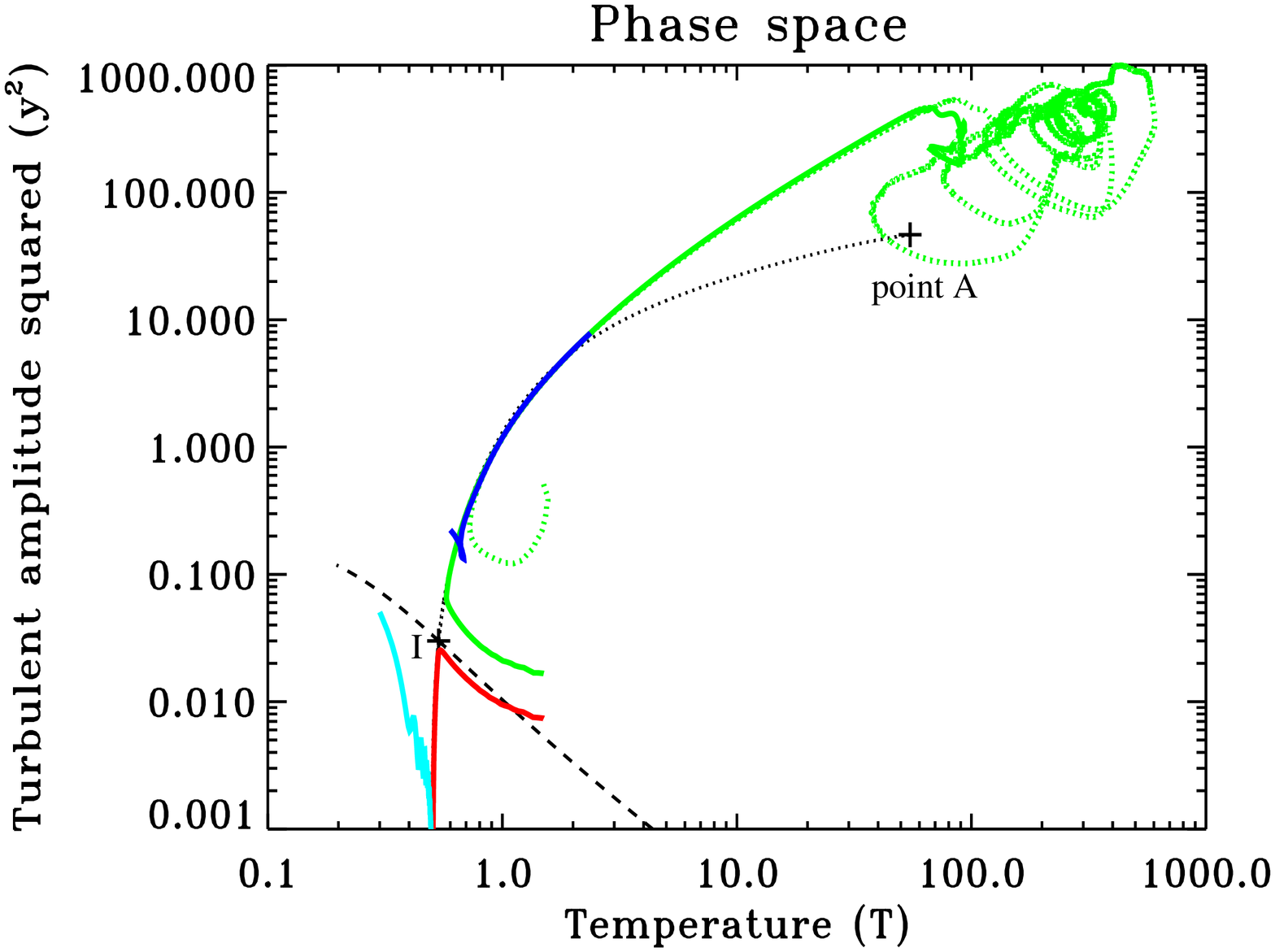,width=6.43cm,angle=-90}
\label{bigphase}
} 
\caption{Comparison with simulations of cubic domains. The separatrices of Fig. 1 are overlayed with five test trajectories $(\bT,\by)$
  in our simulations. The dotted green trajectory is the case `Ct'
  simulation. 
}
\label{phase}
\end{figure}
 
\subsection{The route towards saturation}

 Here we focus on a typical cubic domain simulation, namely, the case `Ct'
 simulation (see appendix \ref{params} for the nomenclature: case `Ci' is isothermal, case `Cc' has cooling, case `Ct' has cooling and a temperature-dependent resistivity). But
 these comments are also valid for other cubic domain cases `Ci' and
 `Cc'.  From the initial conditions, the fastest growing mode is a
 vertical 2-channel flow which soon dominates. The mode grows in amplitude on a time scale no shorter than
 the orbital time scale. Meanwhile, the Alfvén speed increases and its
 crossing time through the box gets shorter and shorter.  The total
 pressure balance across the simulation box is hence better and better
 realised. Therefore the growth of the 2-channel proceeds with a
 nearly uniform total pressure. The location of the two counter
 flowing channels corresponds to the nulls of magnetic
 pressure. Consequently, the thermal pressure must be maximal within
 the channels, and that pressure must increase as the amplitude of the
 mode rises. The result is a sharp increase in density at the location
 of the channels which get thinner and thinner due to mass
 conservation.

 On either side of these channels, the magnetic field is almost uniform
 with same direction and opposite sign. The channels are also the current
 sheets where these two ordered fields reconnect. Ohmic heating is
 thus maximal at the location of these channels, which experience the
 highest temperature in the computational domain for non isothermal
 simulations (case `c' or `t').
  
\begin{figure}
\centering
\subfigure[Initial conditions (after first transient evolution)]{
\label{Ct.0}
\includegraphics[width=0.25\textwidth]{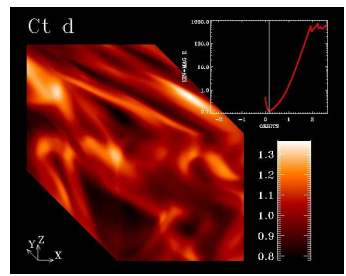}}
\subfigure[The growing channel]{
\label{Ct.1}
\includegraphics[width=0.25\textwidth]{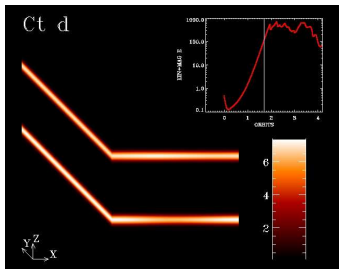}}
\subfigure[The steady state]{
\label{Ct.2}
\includegraphics[width=0.25\textwidth]{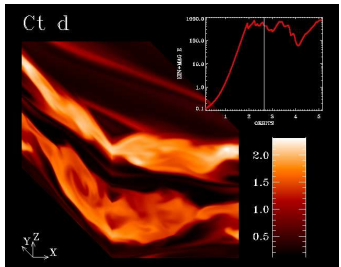}}
\subfigure[At a trough]{
\label{Ct.3}
\includegraphics[width=0.25\textwidth]{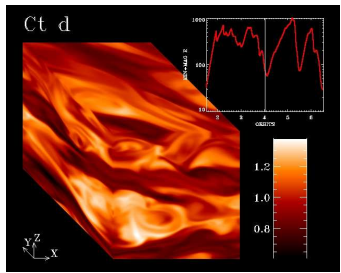}}
\label{Ct}
\caption{Several density slices at various times of the simulation in
  case `Ct'. The evolution of the total kinetic plus magnetic energy ($\by^2$) in the perturbation is also given in the top right hand corner of each figure.}
\end{figure}

 Fig. \ref{Ct.1} illustrates the simulation `Ct' during this phase
 of channel growth. The surprising sharpness of the channel feature is due
 both to the cube geometry of the box and to the linear filtering efficiency
 of case `t'. Indeed, when the resistivity increases (for lower temperatures) towards the
 marginal stability, fewer and fewer modes are available until only
 one mode is unstable. Strong channel features are therefore likely to
 be a characteristic of flows with a temperature close to the critical
 `I' point.  Also, the cubic geometry of the box does not allow for
 unstable parasitic modes as computed by \cite{GX94} to fit in the
 computational domain, so that the growing channel mode remains
 unhindered until very high amplitudes. As  detailed in \cite{LLB09} (for isothermal gases), it is then the
 shrinking of the channel width which decreases the wavelength of unstable
 parasites so that they finally fit in the box and start attacking the
 channel flow.

 The growing process of the channel flow can also be viewed as the
 progressive spatial (as opposed to temporal as in the phase
 portrait)
 separation of the system into two phases: a high
 temperature region of dense and fast moving fluid with a disordered
 magnetic field (within the channels), and a lower temperature region
 of dilute quiet fluid with a strong ordered magnetic field (in
 between the channels). This is of course reminiscent of the phase
 separation described by our reduced model into an active and a quiet
 region. However, the two states are now coupled. For example, the
 fluid in the final quieter region does  transport some
 angular momentum,  as in the work of
 \cite{FS03} where a dead zone is activated by its neighbouring active
 zone. It also has a temperature significantly higher than the quiet state temperature $T_0$ due
 both to thermal diffusion and to the dissipation of the magnetic
 field.
 The partitioning of the medium into a dense and dilute phase, while
 the total pressure is kept uniform,
 resembles the classical Field thermal instability:
  high and low density regions appear at uniform thermal
 pressure. We recall that \cite{BL08} interpreted the stability around
 the point `I' thanks to a criterion similar to the Field criterion. The
 present study provides further evidence for this view.
 
\subsection{The saturated state} 
Even after the onset of the parasitic modes, both channels remain
relatively well defined: we find that cubic domains lead to permanent
channel flows \citep[as witnessed in][]{LLB09}. The parasites soon
saturate, and the two channels remain flowing with travelling
wave-like structures (as in saturated kink modes)  or occasional plasmoids
(saturated pinch modes) as illustrated on Fig. \ref{Ct.2}. These
structures are reminiscent of the saturated kink and pinch modes that
feed on magnetised jets as observed in \cite{B98}.  The parasites' perturbation
of one channel are sometimes strong enough to diffuse and meet the
parasites of the other channel, at which point some stronger coupling
occurs, the channel flow slows and becomes more disordered, although
two blocks with opposite velocities can still readily be identified as
seen on Fig. \ref{Ct.3}.  These events correspond to sudden dips
in the evolution of the total energy contained in the computational
domain. After a short while, two well defined channels reform, with
their saturated parasites. 

  Simulations with shorter cooling time scales have less pressure
  support within the channel flows. They show thinner channels
  which break down sooner and yield lower saturation levels.

\section{Other geometries}
\label{geom}
We have thus far limited the discussion  to
cubic domains, as these seem particularly fit for our reduced
model. We briefly report here some of the potentially interesting behaviour we
encountered in other geometries.

\subsection{Bar results}
  Many MRI simulations have been performed in computational domains
  elongated in the azimuthal direction
  \citep{HGB95,F00,SI01,F07I,F07II,LL07}. Here we focus on
  the results we obtained in a similar geometry (our case `B'
  with $L_x=1$ and $L_y=4$).

  The bar geometry still does not allow for any \cite{GX94} parasites
  to fit in the computational domain for the channel flow of maximum
  growth.  Indeed, the parasites' wave vectors are restricted to a
  sector of angle less than $\pi/2$ encompassing the direction of the
  channel \citep{GX94,LLB09}.  The direction of the dominant channel
  flow is close to $\pi/4$ with respect to the direction axes. Hence
  the $x$-direction wavelength of a parasite needs to be bigger than
  the vertical extent of the channel in order to grow.  As the channel
  gets thinner, however, compressible parasites may fit in the
  computational domain and attack the structure \citep{LLB09}. 
 
\begin{figure}
\centerline{
\begin{tabular}{c}
\psfig{file=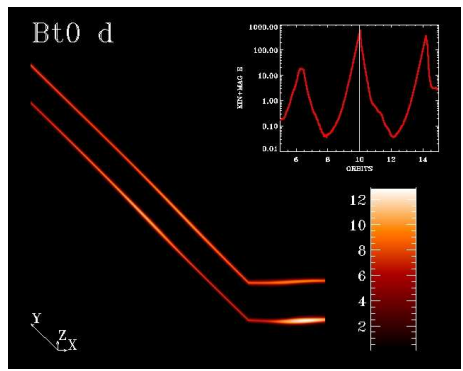,width=6cm}\\
\end{tabular}
} 
\caption{Density slices for case `Bt0' on its way near the tip of a peak in total energy.
 }
\label{Bt}
\end{figure}

The
  chief difference between the cube and the bar is that in the latter
  there is enough room in the $y$-direction for the parasites to
  divert one channel into the path of the other, an event that leads
  to a dramatic collapse of the whole channel structure. Non-linear
  interactions quickly render the flow isotropic while reconnection
  brings the perturbation back to linear amplitudes. Depending on the
  overall resistivity (hence on the overall temperature), a new
  channel mode can or cannot grow. Depending on its thermal inertia,
  the system either enters a new cycle of eruption or decays. We can
  get immediate decay (case `Bt'), a few mild eruptions before decay
  (case `Bt' with less initial seed for parasites), infinite sequence
  of mild eruptions (case `Bt' with a longer cooling timescale),
  infinite sequence of moderate eruptions (case `Bi' and `Bc') or
  infinite sequence of strong eruptions (case `Bt0', see Fig. \ref{Bt}). In case `Bt0', the
  resistivity function is designed such that even for the radiative balance
  temperature there is still one unstable vertical MRI mode.  In the
  next section, we devise a new reduced model which accounts for and
  sheds light on this wide variety of behaviour.

\subsection{Eruptive reduced model}
The previous discussion emphasises the role of the parasitic modes for
the main growing mode which usually is a 2-channel mode. These
parasites were described in \cite{GX94} and \cite{LLB09} as a
horizontal perturbation on top of a vertical main mode of evolution.
Here we separate each field variable $X$ into its horizontally
averaged contribution $\langle\, X\,\rangle _z$ (which only depends on the
altitude $z$) and its remainder $\delta X=X-\langle\, X\,\rangle _z$.  The
component $\langle\, X\,\rangle _z$ naturally traces the main vertical
`channel' when one is dominating. And the component $\delta X$ can be
thought of the `perturbations' on top of this average profile.

\begin{figure}
\centerline{
\begin{tabular}{c}
\psfig{file=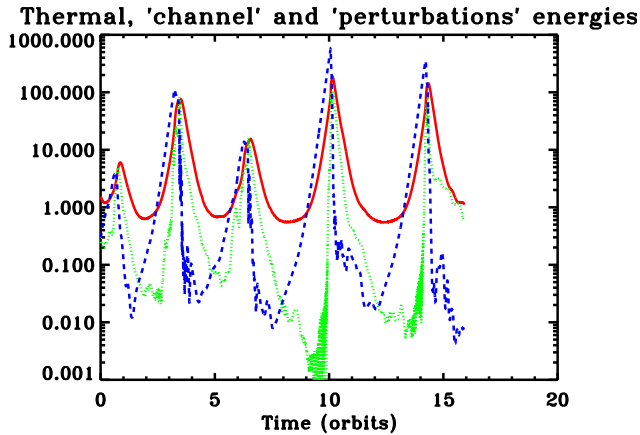,width=6cm,angle=+90}\\
\end{tabular}
} 
\caption{ Evolution of variables $\bT$ (red, solid), $\byc^2$ (blue,
  dashed) and $\byp^2$ (green, dotted) (see text for definitions) in
  the case `Bt0' simulation. 
}
\label{b2}
\end{figure}

We now look at the separate evolution of the energies contained in
each of the components.  In particular, we monitor the magnetic energy
in the `channel' component \beq \byc^2=\frac1{4\pi}\langle\, \langle\, \Bb\,\rangle
_z\bdot\langle\, \Bb\,\rangle _z\,\rangle \eeq and in the `perturbations' component
\beq \byp^2=\frac1{4\pi}\langle\, b^2-\langle\, \Bb\,\rangle _z\bdot\langle\, \Bb\,\rangle
_z\,\rangle \eeq which are displayed on Fig. \ref{b2} along with the
quantity $\bT=\langle\, e\,\rangle $. We set the beginning of an eruptive
cycle at a minimum of the `channel' energy, where a dominant channel
mode is just starting to emerge and grow. At this point
`perturbations'  are still decaying roughly at an orbital rate and they
dominate the total energy: the medium is vertically homogeneous in a
statistical sense. The perturbations start to grow only after the
channel amplitude has reached a certain threshold around 1 in our
units.  Then on the growth rate of the perturbations keeps
increasing, in agreement with the idea that the growth of parasites is
proportional to the channel amplitude \citep{GX94,LLB09}. When the
perturbations reach an amplitude comparable to the channel, the
channel collapses and the medium is perturbation dominated. At
this point, the system is in a state of decaying turbulence
superimposed with spiral density waves, until the amplitudes become
linear again, and a new channel mode emerges.

\begin{figure}
\centerline{
\begin{tabular}{c}
\psfig{file=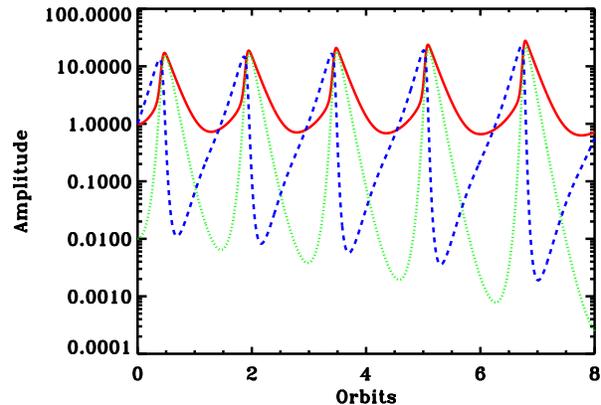,width=6cm,angle=+90}\\
\end{tabular}
} 
\caption{Evolution of variables $T$, $\yc^2$ and $\yp^2$ for the
  reduced equations (\ref{red3})-(\ref{red4}) with $A=0.3$,
  $W=3/2$. The initial values are $T=1$, $\yc=1$ and $\yp=0.1$. 
}
\label{red}
\end{figure}

The description above naturally leads to a simple set of reduced
equations where our former variable $y$ is now detailed into two
components $\yc$ and $\yp$: 
\beq \label{red3}
\dd{\yc}{t}=\sigma(T)\yc-A\yc^2\yp \mbox{,}
\eeq
\beq
\dd{\yp}{t}=(\yc-1)\yp 
\eeq
and
\beq
\dd{T}{t}=W\yc^2-\dd{\yc^2}{t}-\Lambda(T) \mbox{.}
\label{red4}
\eeq 
The former non-linear saturation term is now replaced by a
coupling term between the channel and the perturbations. The square
exponent for $\yc$ in $A\yc^2\yp$ accounts for the observation that
the breakup of a channel arises when two parasites interact. This
simple reduced model is able to reproduce qualitatively each of the
behaviours experienced in the simulations of the previous section,
provided it is started with the right set of initial conditions for
the channel and perturbation component (see Fig. \ref{red} for case
`Bt0'). 

The main missing piece in this reduced model is how the
decaying part of the cycle determines the starting amplitudes for the
next cycle. The dice are rolled again during each phase of decaying
turbulence. 

\subsection{Slab results}
\begin{figure}
\centerline{
\begin{tabular}{c}
\psfig{file=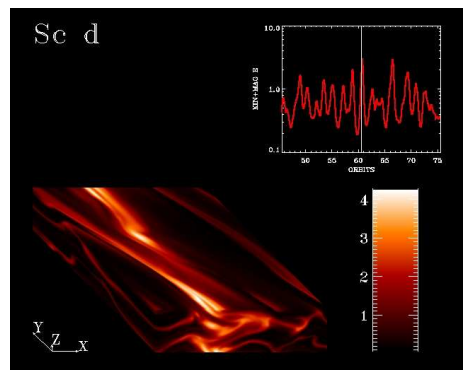,width=6cm}\\
\end{tabular}
} 
\caption{Density slices for the case `Sc' simulation at a point where
  a channel-like structure stands out.
}
\label{slab}
\end{figure}
  We also tested geometries with a slab aspect ratio $L_x=L_y=4$ while
  $L_z=1$.  In this case, \cite{GX94} parasites can fit in the
  computational domain, but the amplitude of the resulting channels
  remain small so that a parasite analysis cannot be applied
  \citep[see][]{LLB09}.  Case `St' decays from the beginning, as for
  case `Bt', but case `St0' (with $\eta_1=0.007$) is similar to case
  `Sc': intermittently, at the peak of the magnetic activity, a kinky
  channel-like structure (see Fig. \ref{slab}) stands out. It washes
  out very quickly. This agrees with what \cite{B08} observe in their
  aspect ratio study.

The flow cannot be described as accurately with a simple model, as
many modes of order 1 amplitude are present, all interacting with one
another. However, the reduced model of equations (\ref{red2a}) and
(\ref{red2b}) may still be used for qualitative comparison, but with
a higher value for the constant $A$.

\section{Summary and prospects.}
  We have performed Cartesian shearing-box simulations in conditions near
  marginal stability, which exist at the border between dead and
  active zones in accretion disks. The linear relative filtering at low
  amplitudes helps select prominent channel flows. This makes the
  system suitable for a description with  simple reduced models
  which are able to account for general properties of the simulations
  such as internal energy, mechanical energy and momentum transport.

We  investigated the influence of the thermal properties of the gas on
  the outcome of our simulations.
  The aspect ratio of our computational domain also has a very strong
  impact on their outcome. This is very likely due to the non-linear dynamics of
  interactions between various MRI modes \citep[see][for
    details]{LLB09}. 

 Stratified and global simulations of disks are
  now required to determine which geometrical configuration is selected
  in real disks. Also, the presence of a mean vertical magnetic field
  in our simulations sets a preferred scale for the channel:
  simulations without zero net mean field still need to be
  investigated.

 Last, we implied a dependence of the resistivity on the
 temperature which originally is mainly due to the electron
 fraction. The time scales for the recombination of electrons can be
 large and time-dependent chemistry has then to be accounted for. This
 eventually has to be included in order to link with the work of
 \cite{IN08}, for example.
%, but it should not in principle be very different from the
% temperature evolution.
\section*{Acknowledgements}
Many thanks to S. Fromang for providing us with his version of the
Zeus3D code, from which we built the version used in this paper (find
it on http://magnet.ens.fr). We acknowledge the invaluable help of
J.-F. Rabasse for the efficiency of his management of the SGI Altix
450 machine, on which all the simulations presented in this paper were
performed. This work as well as half of the machine was supported by a
Chaire d'Excellence awarded by the French ministry of Higher Education
to S. Balbus. The other half of the machine was subsidised by the
region \^Ile de France.  Thanks to the French Embassy in the United Kingdom, PL also benefited from a french government fellowship at Churchill college (Cambridge) where part of this work was assembled.  Thanks to A. Ciardi for pointing out a
problem with \small{ZEUS3D} on the equilibrium solution.  Thanks to G. Lesur for the suggestion that
we include parasites in the reduced model.

 \appendix
\section{Numerical Setup}
\label{setup}

\subsection{Numerical method}
Our code is basically the \small{ZEUS3D} code \citep{ZI,ZII} with the shearing-box implemented as in \cite{HGB95}, and we
started from the version used by S. Fromang \citep{F07I} but amended it
to better suit the problem at hand. We slightly improved the 
parallelisation scheme to increase its efficiency and accuracy.

 We reset the mean magnetic field and the mean momentum to their
initial values every 150 time steps, as in \cite{S08}. This prevents
truncation errors for the mean momentum and magnetic fields from
wandering too far from zero.

  We split the momentum transport step for the Euler equation.  We
first transport the perturbed velocity $\uu$.  This increases the
accuracy of the Van-Leer slopes.  Then we combine the advection of the
mean shear with the Coriolis and tidal forces in an exact analytic
rotation as in Section 3.3 of \cite{GZ07}. This increases the accuracy
of the treatment of epicylic motions.

  We do not use the artificial viscosity of \small{ZEUS3D}. We rather
implement a physical viscous term which takes the same form as in
equation (\ref{Euler}). We rely on this term to degrade kinetic energy
into heat and ensure total energy conservation (see below in Section \ref{params}). 
We rewrite the resistive term in order to be able to use a non uniform
resistivity $\eta$. The resistive criterion for the time-step control
is made non-uniform accordingly. Similarly, note that viscosity is not
assumed uniform in the present implementation of the viscous term,
although we use it as a constant in the present study.

  The diffusion step is done at the end of the standard \small{ZEUS3D} step
and assumes the density is kept fixed.  The cooling term follows and
also assumes the density does not change. The dissipative source heat
terms $\epsilon_{\rm d}$ (viscous and Ohmic) are computed along with
the resistive and viscous terms and added as a constant to the cooling
function $\Lambda_{\rm t}=\Lambda+\epsilon_{\rm d} $. We adopt a
linear cooling function $\Lambda(T)=aT+b$ which allows us to compute
the isochore evolution of the internal energy analytically. Time-step
controls for the thermal diffusion and cooling are based on the
diffusion time across a pixel $\Delta x^2/\chi$ (where $\Delta x$ is
the size of a pixel) and the local cooling time-scale
$\dd{\Lambda}{e}^{-1}$. These controls are never reached in
practise (the limiting time is generally the Courant-Levy-Friedrich
condition in the hottest pixel).

The version of the code used in this paper is available for download
on the MAGNET website http://magnet.ens.fr.
 
\subsection{Units}
  We use the orbital timescale $1/\Omega$ as our unit time. We use the
  vertical extent $L_z$ of the computational domain as our unit
  length.  We use the mass initially contained in a cube of size $L_z$
  as our unit mass, so that the initial average density in the box is
  $\langle\,\rho\,\rangle=1$ (the brackets denote volume averages over
  the full box).  In addition, most of our runs begin with $\langle\, p
  \,\rangle=1$ in our units.  The initial pressure scale height is
  therefore $H=1/\sqrt{\gamma}\simeq0.77 L_z$: the unit length is of
  the order of the initial pressure scale height. The scale height
  later varies as the average temperature in the box changes.

\subsection{Thermal properties}
\label{thermal}
 In the isothermal
case, the pressure is simply related to the density by $p=c_0^2 \rho$
where $c_0$ is the initial sound speed in the non-isothermal runs.
For the non-isothermal cases, the cooling function $\Lambda$ is a function of the temperature which we take to be linear
$$\Lambda(T)=aT+b.$$ In effect, this can be understood as the first order
Taylor expansion of a more realistic cooling function around an intermediate
 temperature between the active and quiet states mentioned in the introduction.

 The steady-state temperature $T_0$ results from the balance
between viscous heating and radiative cooling. It may thus be written
as \beq T_0= (\nu \frac{q^2\Omega^2}4-b)/a=T_r+\nu
\frac{q^2\Omega^2}{4a} \eeq which is slightly higher than the radiative
equilibrium temperature $T_r=-b/a$ due to viscous heating.
We fix the radiative equilibrium temperature to $T_r=1/2$ and hence
adopt \beq \Lambda(T)=a(T-\frac12).  \eeq 

The cooling time scale then depends only on the parameter $a$, which
we set to $a=1$ in most of our runs. Larger values for $a$ yield
similar results as in the isothermal case. Smaller values mean a
longer cooling time $1/a$ and consequently a statistically stationary
thermal state is reached after a longer integration time. Small values
for $a$ also yield higher equilibrium temperatures which constrain the
time-step to much lower values due to the Courant-Friedrich-Levy (CFL)
condition.

\subsection{Initial conditions}
\label{IC}
\begin{figure}
\centerline{
\begin{tabular}{c}
\psfig{file=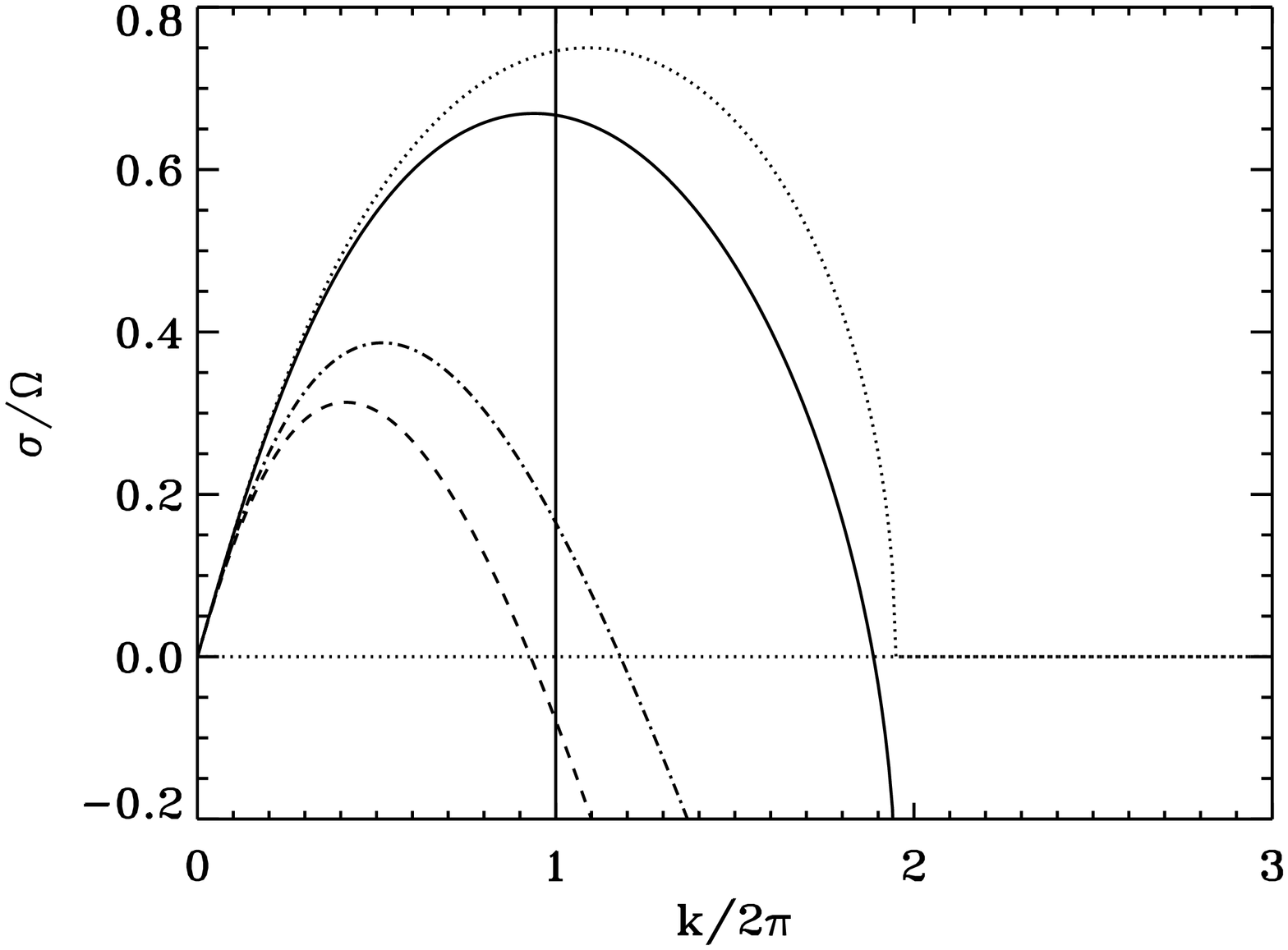,width=6cm,angle=+90}\\
\end{tabular}
} 
\caption{ MRI dispersion relation for the vertical wave numbers in the
  conditions of the main simulations of this paper. Viscosity and
  resistivity are included. We display the real part of the fastest
  growing eigenvalue in units of the orbital time scale. The actual
  dispersion relation can be found in Lesaffre \& Balbus (2007)
% \cite{LB07}
, equation (16). The
  dotted line is for no ideal MHD. The solid line is for case `c'
  ($\eta=\nu=2\times 10^{-3}$).  The dashed line is for case `t' with 
  $T=0.5$ (radiative equilibrium) and the dash-dotted line is for
  $T=2/3$ (as in the initial conditions, which have $\langle\, p\,\rangle =1$). The
  vertical line indicates the vertical wavenumber of the box. The
  wavenumbers resolved by our simulations are right of this line, up
  to $ k = 64 \pi $ for our standard resolution.  
}
\label{disp}
\end{figure}

  The mean vertical magnetic field is set to the value
  $B_0=\sqrt{2/400\pi}$ which corresponds to a plasma $\beta$ value of
   around
  $10^2$
 when the average pressure is $\langle\, p\,\rangle =1$. For
  that value of the magnetic field, the wavelength of the MRI mode of
  maximum growth rate just fits in the vertical extent of the box (see
  Fig. \ref{disp}).

  Because the system supports two stationary states accessed by
different initial conditions, we must then be careful how we
characterise these.  The initial configuration consists of a
perturbation added to the stationary flow described above.

   Random perturbations with low initial amplitude such as
used by \cite{HGB95} lead to a long initial transient phase during
which MRI modes grow and other components of the perturbation field
decay, until the fastest growing MRI modes stand out. During this
linear filtering phase, the temperature as well as the magnitude of
the perturbation may change significantly. 

  In order to shorten this phase, we prepare the system in a state
  that is already linearly filtered to some extent. We choose to build
  the initial perturbation with a combination of some of the fastest
  growing modes and add a random seed.  That mixture of a random
  perturbation and a selection of growing modes considerably
  shortens the initial transient phase.  With this setup, we are
  indeed able to eliminate the initial strong first channel flow
  which appears in the time evolution of MRI shearing-box simulations.

  We set the initial perturbation in the Fourier space of a larger
domain and with a small scale cut off at intermediate scales. We then
select only those modes which fit in the box. That way the same
perturbation is used for runs with different geometries, different
resolutions or different ways of sharing the domain between the
processors units. To work in the Fourier space also allows to properly
clean the divergence of the initial magnetic field.
  
  Finally, we normalise the initial perturbation such that the maximum
radial velocity is sonic (in fact, equal to 1 in our units). That
high initial amplitude maximises our chances to first pick up the
active state.

\subsection{Parameter study}
\label{params}
  The main parameter we varied was the aspect
  ratio of the computational domain. The vertical size of the box
  $L_z$ was kept fixed to $L_z=1$. We set its radial size $L_x$ and
  azimuthal size $L_y$ to $L_x=L_y=1$ for the cube (first letter `C'),
  to $L_x=1$ and $L_y=4$ for a more classical azimuthal bar (first
  letter `B') aspect ratio, and to $L_x=L_y=4$ for a slab (first
  letter `S') aspect ratio.  

  We also changed the microphysical properties of the gas. We tested
  isothermal runs (second letter `i'), runs with constant resistivity
  and cooling (second letter `c') and runs with a varying resistivity
  and cooling (second letter `t'). Some runs were performed with a
  lower resistivity and have a trailing zero in their coding name
  (simulation `Bt0', for example, is a bar simulation with a varying
  resistivity lower than that of simulation `Bt').

\subsection{Resolution}
 We use an isotropic resolution as advised by \cite{LB07}. Our pixels
  are cubes of side length $\Delta x=1/64$ for our standard
  resolution. We halved that value for the case `t' of each of the
  geometries for a resolution test on shorter integration times. The
  size of our simulations was then, in terms of zone numbers: $64^3$
  for the cubes, $64\times 256\times64$ for the bars and $256\times
  256\times64$ for the slabs (and eight times these numbers for the
  higher resolution tests).  Our value for the Courant
  number \citep{LB07} ranges from 0.2 for cases `B' and `S' to 0.5 for
  case `C'. A low Courant number is necessary especially for case `t'
  where large resistivity values can be obtained because our scheme
  uses an explicit resistive term. We aimed at fulfilling an
  integration time of a hundred of orbits. However, that time span
  could not be reached in case `C'. Case `Bt0' is run for up to 330
  orbits. In general, our simulations are run for time-step numbers
  of the order of a million up to six millions.

\subsection{Dissipation and diffusion coefficients}
\label{dissipation}

Total energy is best conserved when the physical dissipation is much
higher than the numerical dissipation. We use equation (50) of
\cite{LB07} with $\Delta x=1/64$, $\beta=100$, $C=0.5$ and $k=64 \pi$
to get an upper bound estimate of the total numerical dissipation
coefficient $\eta_{\rm N}+\nu_{\rm N} < 5\times 10^{-3}$. Accordingly, we
choose our minimal physical dissipation coefficients as
$\eta_0=\nu=2\times 10^{-3}$. Our standard runs are hence only
marginally resolved in places where strong gradients occur or where
the magnetic field becomes very large. Our control runs with twice the
resolution show that this does not impact our results except in case
`C', where we show only the results at a resolution of $128^3$. According to 
the same line of thought, we include thermal diffusion with
$\chi=2\times 10^{-3}$ in cases `c' and `t' in order to avoid spurious
thermal diffusion effects.

  The resistivity in cases `i' and `c' is simply set to
  $\eta=\eta_0$.  In case `t', we adopted a resistivity of the form
  \beq
\label{etat}
\eta(T)=\eta_0+\eta_1 T^{-3/2}\eeq to mimic the power law dependence of
  the resistivity in a fully ionised plasma. The purpose was to test
  the validity of our simple dynamical system: both an active and a
  quiet state should be within reach of the system. Hence, the
  coefficient $\eta_1$ is chosen reasonably small (to avoid putting
  too much constraint on the time step due to high values of
  resistivity), while forcing all the MRI modes that fit in the
  computational to decay when the temperature is $T=0.5$. We
  use the values $\eta_1=1.2\times 10^{-2}$ for standard runs and
  $\eta_1=7\times 10^{-3}$ for lower resistivity runs (case `Bt0', for
  example). The MRI dispersion relations for various
  realisations of the dissipation coefficients are plotted in Fig.
  \ref{disp}.

\bibliographystyle{mn}
\bibliography{biblio}
\end{document}